# Octonionic Hyper-strong and Hyper-weak Fields and Their Quantum Equations

Zihua Weng

*School of Physics and Mechanical & Electrical Engineering, P. O. Box 310, Xiamen University, Xiamen 361005, China*

**Abstract**

Based on Maxwellian quaternionic electromagnetic theory, the octonionic hyper-strong field, hyper-weak field and the quantum interplays can be presented by analogy with octonionic electromagnetic, gravitational, strong and weak interactions. In the hyper-weak field, the study deduces some conclusions of field source particles and intermediate particles which are consistent with Dirac equation, Yang-Mills equation, Schrodinger equation and Klein-Gordon equation. It predicts some new intermediate particles and observed abnormal phenomena about galaxy clusters which may be caused by the hyper-weak field. In the hyper-strong field, the paper draws some conclusions of the field source particles and intermediate particles which are coincident with Dirac-like equations. It predicts some new intermediate particles, three kinds of colors, and some observed abnormal phenomena which may be caused by the sub-quarks in hyper-strong field. The researches results show that there may exist the some new field source particles in the hyper-strong and hyper-weak fields.

## 1. Introduction

Nowadays, there still exist some movement phenomena of galaxy clusters which can't be explained by current gravitational theory even the hypothesis of dark matter. Therefore, some scientists doubt the universality of current gravitational theory, and then bring forward the hyper-weak field theory to explain the abnormal phenomena of galaxy clusters etc.

Meanwhile, the physics of strong interactions is undoubtedly one of the most challenging areas of modern science. The strong interactions keep providing new experimental observations, which were not predicted by 'effective' theories. It retains the problems in describing all the observed phenomena simultaneously. And then some scientists bring forward the hyper-strong field theory to explain the phenomena of sub-quarks and the associated problem etc.

---

A new discernment on the problem of hyper-strong and hyper-weak fields can be given by the concept of the octonionic space. The octonionic space has been applied to electromagnetic-gravitational field and strong-weak field, and would be extended similarly to the hyper-strong and hyper-weak fields. All these fields are related to 'SpaceTime Equality Postulation' [1]. The postulation can be explained further as: (1) the spacetime is the extension of fundamental interaction characteristics, and there is no such spacetime without any fundamental interaction, (2) each fundamental interaction possesses its unique spacetime, and all these spacetimes are equal, (3) the spacetime of each fundamental interaction is quaternion space.

According to the 'SpaceTime Equality Postulation', the hyper-strong field and hyper-weak field can be described by the octonionic space. By analogy with previous research results of electromagnetic-gravitational field etc., the paper describes the quantum properties of the hyper-strong field, and draws some conclusions and predictions associated with the quantum feature of sub-quarks. Its some new and unknown particles and energy can be derived from the octonionic hyper-strong field theory. The study develops the quantum theory of the hyper-weak field, and draws some conclusions and predictions associated with the quantum characteristics of the hyper-weak field. Its some new and unknown particles can be used to explain the abnormal movement phenomena of galaxy clusters etc. [2, 3]

**2. Octonionic electromagnetic-gravitational field**

The electromagnetic interaction theory can be described with the quaternion algebra. In the treatise on electromagnetic theory, the quaternion algebra was first used by J. C. Maxwell to describe the equations set and various properties of the electromagnetic field. The spacetime, which is associated with electromagnetic interaction and possesses the physics content, is adopted by the quaternionic space. According to the 'SpaceTime Equality Postulation', the gravitational interaction can be described by the quaternion likewise.

The electromagnetic and gravitational interactions are interconnected, unified and equal. Both of them can be described in the quaternionic space. Based on the conception of the space expansion etc., two types of quaternionic spaces can combine into an octonionic space (Octonionic space E-G, for short). In the octonionic space, various characteristics of electromagnetic and gravitational interactions can be described uniformly, and some equations set of the electromagnetic-gravitational field can be attained.

2.1 Equations set of electromagnetic-gravitational field

There exist four types of subfields and their field sources in the electromagnetic-gravitational field. In Table 1, the electromagnetic-electromagnetic subfield is 'extended electromagnetic field', and its general charge is E electric-charge. The gravitational-gravitational subfield is 'modified gravitational filed', and its general charge is G gravitational-charge. Meanwhile, electromagnetic-gravitational and gravitational-electromagnetic subfields are both long range fields and candidates of the 'dark matter field'. And their general charges (G electric-charge and E gravitational-charge) are candidates of 'dark matter'. The physical features of the dark matter meet the requirement of field equations set in Table 2.

Table 1. Subfield types of electromagnetic-gravitational field

|  | Electromagnetic Interaction | Gravitational Interaction |
|---|---|---|
| electromagnetic quaternionic space | electromagnetic-electromagnetic subfield, E electric-charge, intermediate particle $\gamma_{ee}$ | gravitational-electromagnetic subfield, E gravitational-charge, intermediate particle $\gamma_{ge}$ |
| gravitational quaternionic space | electromagnetic-gravitational subfield, G electric-charge, intermediate particle $\gamma_{eg}$ | gravitational-gravitational subfield, G gravitational-charge, intermediate particle $\gamma_{gg}$ |

The particles of the ordinary matter (electron etc.) possess the E charge together with G gravitational-charge. The particles of dark matter may possess the G electric-charge with E gravitational-charge, or G gravitational-charge with G electric-charge, etc. Where, $k = k^B{}_{E\text{-}G} = c$ is the constant.

The electromagnetic-electromagnetic subfield is the extended electromagnetic interaction. The variation of field energy density has direct effect on the force of charge and current. As shown in Aharonov-Bohm experiment, field potentials have effects on the potential energy etc. And the change of field potential can impact the movement of field sources.

Table 2. Equations set of electromagnetic-gravitational field

| Spacetime | Octonionic space E-G |
|---|---|
| $\mathcal{X}$ physical quantity | $\mathcal{X} = \mathcal{X}_{E\text{-}G}$ |
| Field potential | $\mathcal{A} = \lozenge^* \circ \mathcal{X}$ |
| Field strength | $\mathcal{B} = \lozenge \circ \mathcal{A}$ |
| Field source | $c\mu \mathcal{S} = c\,(\mathcal{B}/c + \lozenge)^* \circ \mathcal{B}$ |
| Force | $\mathcal{Z} = c\,(\mathcal{B}/c + \lozenge) \circ \mathcal{S}$ |
| Angular momentum | $\mathcal{M} = \mathcal{S} \circ (\mathcal{R} + k_{rx}\mathcal{X})$ |
| Energy | $\mathcal{W} = c\,(\mathcal{B}/c + \lozenge)^* \circ \mathcal{M}$ |
| Power | $\mathcal{N} = c\,(\mathcal{B}/c + \lozenge) \circ \mathcal{W}$ |

The gravitational-gravitational subfield is the modified gravitational field, which includes familiar Newtonian gravitational field. The modified gravitational field has the following prediction. The planetary orbits possess the near coplanarity, near circularity and corevolving, and the planets own rotation property. The centrifugal force of celestial body will change and lead to fluctuation of revolution speed, when the field energy density or angular momentum of celestial body along the sense of revolution varied.

The electromagnetic-gravitational and gravitational-electromagnetic subfields are both long range fields and candidates of dark matter field. The field strength of the electromagnetic-gravitational and gravitational-electromagnetic subfields may be equal, and both of them are slightly less than that of gravitational-gravitational subfield. Two types of the field sources possessed by the dark matter field would make the dark matter particles diversiform. The research results explain that some observed abnormal phenomena about celestial bodies are caused by either modified gravitational interaction or dark matter.

2.2 Quantum equations set of electromagnetic-gravitational field

In the octonionic space E-G, the wave functions of the quantum mechanics are the octonionic equations set. And the Dirac and Klein-Gordon equations of the quantum mechanics are actually the wave equations sets which are associated with particles' wave function $\Psi$. Where, the wave function $\Psi = \mathcal{M}/\hbar_{E-G}$ can be written as the exponential form with the octonionic characteristics. The coefficient h is Planck constant, and $\hbar = \hbar_{E-G} = h/2\pi$.

By comparison, we find that the Dirac equation and Klein-Gordon equation can be attained respectively from the energy equation and power quantum equation after substituting operator c ($\mathcal{B}/c + \diamond$) for {$\mathcal{W}/(c\hbar_{E-G}) + \diamond$}. By analogy with the above equations, the Dirac-like and the second Dirac-like equations can be procured from the field source and force quantum equations respectively. Wherein, $\mathcal{B}/\hbar_{E-G}$ is the wave function also.

Table 3.  Quantum equations set of electromagnetic-gravitational field

| Energy quantum | $\mathcal{U} = (\mathcal{W}/c\hbar_{E-G} + \diamond)^* \circ \mathcal{M}$ |
|---|---|
| Power quantum | $\mathcal{L} = (\mathcal{W}/c\hbar_{E-G} + \diamond) \circ \mathcal{U}$ |
| Field source quantum | $\mathcal{T} = (\mathcal{W}/c\hbar_{E-G} + \diamond)^* \circ \mathcal{B}$ |
| Force quantum | $O = (\mathcal{W}/c\hbar_{E-G} + \diamond) \circ \mathcal{T}$ |

In electromagnetic-gravitational field in Table 3, when the energy quantum equation $\mathcal{U} = 0$, the Dirac and Schrodinger equations can be deduced to describe the field source particle with the spin 1/2 (electron etc.). When the power quantum equation $\mathcal{L} = 0$, the Klein-Gordon equation can be attained to explain the field source particle with spin 0. When the field source quantum equation $\mathcal{T} = 0$, the Dirac-like equation can be deduced to describe the intermediate particle with spin 1 (photon etc.). When the force quantum equation $O = 0$, the second Dirac-like equation can be attained to explain the intermediate particle with the spin N. Those equations can conclude the results which are consistent with the quantum property of electromagnetic-electromagnetic subfield in certain cases.

**3. Octonionic strong-weak field**

There is an analogy between the octonionic spacetime of the strong-weak field and that of the electromagnetic-gravitational field. According to the 'SpaceTime Equality Postulation', the spacetime, which is associated with the strong interaction and possesses the physics content, is adopted by the quaternionic space. And the spacetime derived from the weak interaction is supposed to be the quaternionic space also.

The strong and weak interactions are interconnected, unified and equal. Both of them can be described in the quaternionic space. By means of the conception of the space expansion etc., two types of the quaternionic spaces can combine into an octonionic space (Octonionic space S-W, for short). In the octonionic space, various characteristics of the strong and weak interactions can be described uniformly, and some equations set of the strong-weak field can be obtained.

3.1 Equations set of strong-weak field

In strong-weak field, there exist four types of subfields and their field sources. The subfields of strong-weak field are short range. In Table 4, the strong-strong subfield may be 'extended strong field', its general charge is S strong-charge. The weak-weak subfield may be regarded as the 'modified weak field', and its general charge is W weak-charge. And the strong-weak and weak-strong subfields may be regarded as two kinds of new and unknown subfields. Their general charges are W strong-charge and S weak-charge respectively.

It can be predicted that the field strength of the strong-weak and weak-strong subfields may be equal, and both of them are weaker than that of weak-weak subfield. Sometimes, the first two subfields may be regarded as 'familiar weak field' mistakenly. Therefore three kinds of colors have to introduce into the familiar weak interaction theory to distinguish above three types of weaker subfields in the strong-weak field. The physical features of subfields of the strong-weak field meet the requirement of Table 5.

Table 4.   Subfield types of strong-weak field

|  | Strong Interaction | Weak Interaction |
| --- | --- | --- |
| strong quaternionic space | strong-strong subfield<br>S strong-charge<br>intermediate particle $\gamma_{ss}$ | weak-strong subfield<br>S weak-charge<br>intermediate particle $\gamma_{ws}$ |
| weak quaternionic space | strong-weak subfield<br>W strong-charge<br>intermediate particle $\gamma_{sw}$ | weak-weak subfield<br>W weak-charge<br>intermediate particle $\gamma_{ww}$ |

In the strong-weak field, the definition of the field strength is very different from that of the electromagnetic-gravitational field, and thus the succeeding equations and operators should be revised. By analogy with the case of the electromagnetic-gravitational field, the octonionic differential operator $\diamond$ in the strong-weak field needs to be generalized to the new operator ($\mathcal{A}/k + \diamond$). That is because the field potential is much more fundamental than field strength, and the strong-weak field belongs to the short range. Therefore the physical characteristics of the strong-weak field can be researched from many aspects. Where, $k = k^A{}_{S\text{-}W}$ is the constant.

Table 5.   Equations set of strong-weak field

| Spacetime | Octonionic space S-W |
|---|---|
| $\mathcal{X}$ physical quantity | $\mathcal{X} = \mathcal{X}_{S\text{-}W}$ |
| Field potential | $\mathcal{A} = \lozenge^* \circ \mathcal{X}$ |
| Field strength | $\mathcal{B} = (\mathcal{A}/k + \lozenge) \circ \mathcal{A}$ |
| Field source | $k\mu\mathcal{S} = k\,(\mathcal{A}/k + \lozenge)^* \circ \mathcal{B}$ |
| Force | $\mathcal{Z} = k\,(\mathcal{A}/k + \lozenge) \circ \mathcal{S}$ |
| Angular momentum | $\mathcal{M} = \mathcal{S} \circ (\mathcal{R} + k_{rx}\mathcal{X})$ |
| Energy | $\mathcal{W} = k\,(\mathcal{A}/k + \lozenge)^* \circ \mathcal{M}$ |
| Power | $\mathcal{N} = k\,(\mathcal{A}/k + \lozenge) \circ \mathcal{W}$ |

3.2 Quantum equations set of strong-weak field

In the octonionic space S-W, the wave functions of the quantum mechanics are the octonionic equations set. And the Dirac and Klein-Gordon equations of the quantum mechanics are actually the wave equations set which are associated with particles' wave function $\Psi$. Where, the wave function $\Psi = \mathcal{M}/\hbar_{S\text{-}W}$ can be written as the exponential form with the octonionic characteristics. The coefficient b is the Plank-like constant.

By comparison, we find that Dirac and Klein-Gordon equations in strong-weak field can be attained respectively from energy equation and power equation after substituting the operator $k\,(\mathcal{A}/k + \lozenge)$ for $(\mathcal{W}/k\hbar_{S\text{-}W} + \lozenge)$. In the strong-weak field, when the energy quantum equation $\mathcal{U} = 0$, the Dirac and Schrodinger equations can be deduced to describe the field source particle with spin 1/2 (quark and lepton etc.). When the power quantum equation $\mathcal{L} = 0$, the Klein-Gordon equation can be obtained to explain the field source particle with spin 0.

Table 6.   Quantum equations set of strong-weak field

| Energy quantum | $\mathcal{U} = (\mathcal{W}/k\hbar_{S\text{-}W} + \lozenge)^* \circ \mathcal{M}$ |
|---|---|
| Power quantum | $\mathcal{L} = (\mathcal{W}/k\hbar_{S\text{-}W} + \lozenge) \circ \mathcal{U}$ |
| Field strength quantum | $\mathcal{G} = (\mathcal{W}/k\hbar_{S\text{-}W} + \lozenge) \circ \mathcal{A}$ |
| Field source quantum | $\mathcal{T} = (\mathcal{W}/k\hbar_{S\text{-}W} + \lozenge)^* \circ \mathcal{G}$ |
| Force quantum | $\mathcal{O} = (\mathcal{W}/k\hbar_{S\text{-}W} + \lozenge) \circ \mathcal{T}$ |

By analogy with the above equations, three kinds of Dirac-like equations can be concluded from the field strength equation, field source equation and force equation respectively. Wherein, $\mathcal{A}/\hbar_{S\text{-}W}$ is the wave function. In the above strong-weak field, those three sorts of Dirac-like equations can be used to describe the properties of the intermediate particles.

3.3 Compounding particles

In strong-weak field, the intermediate and field source particles can be tabled as following. In Table 7, we can find the intermediate boson, lepton, three colors of quarks, and other kinds of new and unknown particles which may be existed in the nature and expressed in parentheses.

Table 7. Sorts of particles in strong-weak field

|  | S strong-charge | W strong-charge | S weak-charge | W weak-charge |
|---|---|---|---|---|
| S strong-charge | (intermediate particle) | green quark | red quark | blue quark |
| W strong-charge | green quark | (intermediate particle) | (?) | (?) |
| S weak-charge | red quark | (?) | (intermediate particle) | (?) |
| W weak-charge | blue quark | (?) | (?) | lepton, intermediate boson |

In the electromagnetic-gravitational field, the field sources and intermediate particles can be tabled as following. In the Table 8, we shall find the photon, electron, and other kinds of unknown particles which may be existed in the nature. Specially, by analogy with above three colors of quarks in Table 7, it can be predicted that there may are the 'green electron' and 'red electron' in the nature, if the familiar electron is denoted as the 'blue' electron. Some particles may possess more than two sorts of general charges. If the quarks possess the S strong-charge, W weak-charge, E electric-charge and G gravitational-charge, they will take part in strong, weak, electromagnetic and gravitational interactions.

Table 8. Sorts of particles in electromagnetic-gravitational field

|  | E electric-charge | G electric-charge | E gravitational-charge | G gravitational-charge |
|---|---|---|---|---|
| E electric-charge | photon | (green electron, dark matter) | (red electron, dark matter) | (blue) electron |
| G electric-charge | (green electron, dark matter) | (intermediate particle) | (dark matter) | (dark matter) |
| E gravitational-charge | (red electron, dark matter) | (dark matter) | (intermediate particle) | (dark matter) |
| G gravitational-charge | (blue) electron | (dark matter) | (dark matter) | mass, (intermediate particle) |

In strong-strong subfield, we can deduce Yang-Mills equation, Dirac equation, Schrodinger equation and Klein-Gordon equation of quarks and leptons. And it is able to infer Dirac-like and the second Dirac-like equations of intermediate particles among the leptons or quarks etc. It may predict that there are some new energy parts and new unknown particles of quarks, leptons and intermediate particles in the subfields.

In the strong-weak field, we can explain why quarks possess three sorts of colors and take part in four kinds of interactions. There exist two new kinds of subfields (weak-strong subfield, strong-weak subfield), which field strengths may be equivalent to that of the weak-weak subfield. Sometimes, we mix up them with the weak-weak subfield mistakenly. So it needs three sorts of 'colors' to distinguish those 'mixed' subfields in the quark theory.

## 4. Hypothetic fields

Based on electromagnetic-gravitational field and strong-weak field, it can be presumed that there exist two new kinds of unknown fields at least theoretically. One is the hyper-strong field, and the other is the hyper-weak field. The first one is much stronger and shorter than the strong-strong subfield, and the last one is much weaker and longer than the gravitational-gravitational subfield.

In comparisons with the following field property, we present the theory of hyper-strong and hyper-weak fields. In the gravitational-gravitational field, the item ($\mathcal{B}^* \circ \mathcal{B}$) of field source definition possesses the field strength. In the strong-weak field, the item ($\mathcal{A} \circ \mathcal{A}$) of field strength definition possesses the field potential. It is easy to find that field potential is more fundamental than the field strength. So we may speculate on that there exist one kind of much stronger and shorter field (hyper-strong field) with the item ($\mathcal{X}^* \circ \mathcal{X}$) of field potential definition, for the physical quantity $\mathcal{X}$ is more fundamental than the field potential. And there exist one kind of much weaker and longer field (hyper-weak field) with the item ($S \circ S$) of the force definition, for the field source $S$ is not as fundamental as the field strength $\mathcal{B}$.

Table 9.  Summary of field definitions and equations

| Field | Hyper-strong field | Strong-weak field | Electromagnetic-gravitational field | Hyper-weak field |
|---|---|---|---|---|
| Intensity | stronger | strong | weak | weaker |
| Field range | hyper-short | short | Long | hyper-long |
| Spacetime | Octonionic space H-S | Octonionic space S-W | Octonionic space E-G | Octonionic space H-W |
| Quantity $\mathcal{X}$ | $\mathcal{X} = \mathcal{X}_{H-S}$ | $\mathcal{X} = \mathcal{X}_{S-W}$ | $\mathcal{X} = \mathcal{X}_{E-G}$ | $\mathcal{X} = \mathcal{X}_{H-W}$ |
| Constant k | $k = k^X_{H-S}$ | $k = k^A_{S-W}$ | $k = k^B_{E-G}$ | $k = k^S_{H-W}$ |
| Operator $\mathcal{Y}$ | $\mathcal{Y} = \mathcal{X}/k + \diamond$ | $\mathcal{Y} = \mathcal{A}/k + \diamond$ | $\mathcal{Y} = \mathcal{B}/k + \diamond$ | $\mathcal{Y} = S/k + \diamond$ |
| Field potential | $\mathcal{A} = \mathcal{Y}^* \circ \mathcal{X}$ | $\mathcal{A} = \diamond^* \circ \mathcal{X}$ | $\mathcal{A} = \diamond^* \circ \mathcal{X}$ | $\mathcal{A} = \diamond^* \circ \mathcal{X}$ |
| Field strength | $\mathcal{B} = \mathcal{Y} \circ \mathcal{A}$ | $\mathcal{B} = \mathcal{Y} \circ \mathcal{A}$ | $\mathcal{B} = \diamond \circ \mathcal{A}$ | $\mathcal{B} = \diamond \circ \mathcal{A}$ |
| Field source | $\mu S = \mathcal{Y}^* \circ \mathcal{B}$ | $\mu S = \mathcal{Y}^* \circ \mathcal{B}$ | $\mu S = \mathcal{Y}^* \circ \mathcal{B}$ | $\mu S = \diamond^* \circ \mathcal{B}$ |
| Force | $Z = k\mathcal{Y} \circ S$ | $Z = k\mathcal{Y} \circ S$ | $Z = k\mathcal{Y} \circ S$ | $Z = k\mathcal{Y} \circ S$ |
| Angular momentum | $\mathcal{M} = S \circ (\mathcal{R} + k_{rx}\mathcal{X})$ | $\mathcal{M} = S \circ (\mathcal{R} + k_{rx}\mathcal{X})$ | $\mathcal{M} = S \circ (\mathcal{R} + k_{rx}\mathcal{X})$ | $\mathcal{M} = S \circ (\mathcal{R} + k_{rx}\mathcal{X})$ |
| Energy | $\mathcal{W} = k\mathcal{Y}^* \circ \mathcal{M}$ | $\mathcal{W} = k\mathcal{Y}^* \circ \mathcal{M}$ | $\mathcal{W} = k\mathcal{Y}^* \circ \mathcal{M}$ | $\mathcal{W} = k\mathcal{Y}^* \circ \mathcal{M}$ |
| Power | $\mathcal{N} = k\mathcal{Y} \circ \mathcal{W}$ | $\mathcal{N} = k\mathcal{Y} \circ \mathcal{W}$ | $\mathcal{N} = k\mathcal{Y} \circ \mathcal{W}$ | $\mathcal{N} = k\mathcal{Y} \circ \mathcal{W}$ |
| Quantity $\hbar$ | $\hbar = \hbar_{H-S}$ | $\hbar = \hbar_{S-W}$ | $\hbar = \hbar_{E-G}$ | $\hbar = \hbar_{H-W}$ |

| Operator $C$ | $C = \mathcal{W}/k\hbar + \diamond$ | $C = \mathcal{W}/k\hbar + \diamond$ | $C = \mathcal{W}/k\hbar + \diamond$ | $C = \mathcal{W}/k\hbar + \diamond$ |
|---|---|---|---|---|
| Energy quantum | $\mathcal{U} = C^* \circ \mathcal{M}$ | $\mathcal{U} = C^* \circ \mathcal{M}$ | $\mathcal{U} = C^* \circ \mathcal{M}$ | $\mathcal{U} = C^* \circ \mathcal{M}$ |
| Power quantum | $\mathcal{L} = C \circ \mathcal{U}$ | $\mathcal{L} = C \circ \mathcal{U}$ | $\mathcal{L} = C \circ \mathcal{U}$ | $\mathcal{L} = C \circ \mathcal{U}$ |
| Field potential quantum | $\mathcal{D} = C^* \circ \mathcal{X}$ | / | / | / |
| Field strength quantum | $\mathcal{G} = C \circ \mathcal{D}$ | $\mathcal{G} = C \circ \mathcal{A}$ | / | / |
| Field source quantum | $\mathcal{T} = C^* \circ \mathcal{G}$ | $\mathcal{T} = C^* \circ \mathcal{G}$ | $\mathcal{T} = C^* \circ \mathcal{B}$ | / |
| Force quantum | $\mathcal{O} = C \circ \mathcal{T}$ | $\mathcal{O} = C \circ \mathcal{T}$ | $\mathcal{O} = C \circ \mathcal{T}$ | $\mathcal{O} = C \circ \mathcal{T}$ |

By analogy with the electromagnetic-gravitational field and strong-weak field, the physical quantity and equations set of the hyper-strong and hyper-weak fields can be deduced in the Table 9. The spacetimes, equations and operators of the last two fields are different to that of the first two fields.

## 5. Octonionic hyper-strong field

There is an analogy between the octonionic spacetime of the hyper-strong field and that of the electromagnetic-gravitational field. According to the 'SpaceTime Equality Postulation', the spacetime, which is associated with the hyper-strong field and possesses the physics content, is adopted by the octonionic space (Octonionic space H-S, for short). In the octonionic space H-S, various characteristics of the hyper-strong field can be described uniformly and some equations set of the hyper-strong field can be obtained.

In the octonionic space H-S, the base $\mathcal{E}$ can be written as
$$\mathcal{E} = (\mathbf{1},\ \vec{i},\ \vec{j},\ \vec{k},\ \vec{e},\ \vec{I},\ \vec{J},\ \vec{K}) \qquad (1)$$

The displacement ($r_0$, $r_1$, $r_2$, $r_3$, $R_0$, $R_1$, $R_2$, $R_3$) in the octonionic space is
$$\mathcal{R} = (r_0 + \vec{i}\, r_1 + \vec{j}\, r_2 + \vec{k}\, r_3) + (\vec{e}\, R_0 + \vec{I}\, R_1 + \vec{J}\, R_2 + \vec{K}\, R_3) \qquad (2)$$
where, $r_0 = ct$, $R_0 = cT$. c is the speed of intermediate particle, t and T denote the time.

The octonionic differential operator $\diamond$ and its conjugate operator $\diamond^*$ are defined as,
$$\diamond = \diamond_{hs} + \diamond_{HS}\ ,\quad \diamond^* = \diamond^*_{hs} + \diamond^*_{HS} \qquad (3)$$
where, $\diamond_{hs} = \partial_{hs0} + \vec{i}\,\partial_{hs1} + \vec{j}\,\partial_{hs2} + \vec{k}\,\partial_{hs3}$, $\diamond_{HS} = \vec{e}\,\partial_{HS0} + \vec{I}\,\partial_{HS1} + \vec{J}\,\partial_{HS2} + \vec{K}\,\partial_{HS3}$. $\partial_{hsj} = \partial/\partial r_j$, $\partial_{HSj} = \partial/\partial R_j$, j = 0, 1, 2, 3.

The octonionic differential operator $\diamond$ meets ($\mathcal{Q}$ is an octonionic physical quantity)
$$\diamond^* \circ (\diamond \circ \mathcal{Q}) = (\diamond^* \circ \diamond) \circ \mathcal{Q} = (\diamond \circ \diamond^*) \circ \mathcal{Q} \qquad (4)$$

In the hyper-strong field, the field potential ($a_0$, $a_1$, $a_2$, $a_3$, $k_{hs} A_0$, $k_{hs} A_1$, $k_{hs} A_2$, $k_{hs} A_3$) is defined as
$$\begin{aligned}\mathcal{A} &= (\mathcal{X}/k + \diamond)^* \circ \mathcal{X} \\ &= (a_0 + \vec{i}\, a_1 + \vec{j}\, a_2 + \vec{k}\, a_3) + k_{hs}(\vec{e}\, A_0 + \vec{I}\, A_1 + \vec{J}\, A_2 + \vec{K}\, A_3)\end{aligned} \qquad (5)$$

where, $\mathcal{X} = \mathcal{X}_{\text{H-S}}$ is the physical quantity in octonionic space H-S; $k_{hs}$ is the coefficient; $k = c = k^X_{\text{H-S}}$ is the constant.

Table 10.  Octonion multiplication table

|   | 1 | $\vec{i}$ | $\vec{j}$ | $\vec{k}$ | $\vec{e}$ | $\vec{I}$ | $\vec{J}$ | $\vec{K}$ |
|---|---|---|---|---|---|---|---|---|
| **1** | 1 | $\vec{i}$ | $\vec{j}$ | $\vec{k}$ | $\vec{e}$ | $\vec{I}$ | $\vec{J}$ | $\vec{K}$ |
| $\vec{i}$ | $\vec{i}$ | **-1** | $\vec{k}$ | $-\vec{j}$ | $\vec{I}$ | $-\vec{e}$ | $-\vec{K}$ | $\vec{J}$ |
| $\vec{j}$ | $\vec{j}$ | $-\vec{k}$ | **-1** | $\vec{i}$ | $\vec{J}$ | $\vec{K}$ | $-\vec{e}$ | $-\vec{I}$ |
| $\vec{k}$ | $\vec{k}$ | $\vec{j}$ | $-\vec{i}$ | **-1** | $\vec{K}$ | $-\vec{J}$ | $\vec{I}$ | $-\vec{e}$ |
| $\vec{e}$ | $\vec{e}$ | $-\vec{I}$ | $-\vec{J}$ | $-\vec{K}$ | **-1** | $\vec{i}$ | $\vec{j}$ | $\vec{k}$ |
| $\vec{I}$ | $\vec{I}$ | $\vec{e}$ | $-\vec{K}$ | $\vec{J}$ | $-\vec{i}$ | **-1** | $-\vec{k}$ | $\vec{j}$ |
| $\vec{J}$ | $\vec{J}$ | $\vec{K}$ | $\vec{e}$ | $-\vec{I}$ | $-\vec{j}$ | $\vec{k}$ | **-1** | $-\vec{i}$ |
| $\vec{K}$ | $\vec{K}$ | $-\vec{J}$ | $\vec{I}$ | $\vec{e}$ | $-\vec{k}$ | $-\vec{j}$ | $\vec{i}$ | **-1** |

5.1 Equations set of hyper-strong field

The hyper-strong field may consist of A and B interactions. In the hyper-strong field, there exist four types of subfields (A-A subfield, A-B subfield, B-A subfield and B-B subfield) and their field sources (A A-charge, A B-charge, B A-charge, B B-charge). The subfields of the hyper-strong field are hyper-short range in Table 11.

Table 11.  Subfield types of hyper-strong field

|  | A Interaction | B Interaction |
|---|---|---|
| A quaternionic space | A-A subfield<br>A A-charge<br>intermediate particle $\gamma_{AA}$ | B-A subfield<br>A B-charge<br>intermediate particle $\gamma_{BA}$ |
| B quaternionic space | A-B subfield<br>B A-charge<br>intermediate particle $\gamma_{AB}$ | B-B subfield<br>B B-charge<br>intermediate particle $\gamma_{BB}$ |

It can be predicted that the field strength of the A-A subfield is stronger than other three subfields. The field strength of the B-A subfield and A-B subfield may be equal, and both of them are weaker than that of B-B subfield. The physical features of each subfield in the hyper-strong field meet the requirement of equations in the Table 12.

In the hyper-strong field, the definition of the field potential is very different to that of other fields (electromagnetic-gravitational or strong-weak fields), and thus the succeeding equations and operators should be revised. By analogy with the case of other fields, the octonionic differential operator $\diamondsuit$ in the hyper-strong field needs to be generalized to the new operator $(\mathcal{X}/k + \diamond)$. That is because the physical quantity $\mathcal{X}$ is much more fundamental than the field potential, and the hyper-strong field belongs to the hyper-short range. So the physical characteristics of hyper-strong field can be studied from many aspects.

The field strength $\mathcal{B}$ of the hyper-strong field can be defined as

$$\mathcal{B} = (\mathcal{X}/k + \diamond) \circ \mathcal{A} \qquad (6)$$

then the field source and the force of the hyper-strong field can be defined respectively as

$$\mu \mathcal{S} = (\mathcal{X}/k + \diamond)^* \circ \mathcal{B} \qquad (7)$$
$$\mathcal{Z} = k (\mathcal{X}/k + \diamond) \circ \mathcal{S} \qquad (8)$$

where, the mark (*) denotes octonionic conjugate. The coefficient $\mu$ is interaction intensity of the hyper-strong field.

The angular momentum of the hyper-strong field can be defined as ($k_{rx}$ is the coefficient)

$$\mathcal{M} = \mathcal{S} \circ (\mathcal{R} + k_{rx} \mathcal{X}) \qquad (9)$$

and the energy and power in the hyper-strong field can be defined respectively as

$$\mathcal{W} = k (\mathcal{X}/k + \diamond)^* \circ \mathcal{M} \qquad (10)$$
$$\mathcal{N} = k (\mathcal{X}/k + \diamond) \circ \mathcal{W} \qquad (11)$$

The above equations show that, the angular momentum $\mathcal{M}$, energy $\mathcal{W}$ and power $\mathcal{N}$ of the hyper-strong field are different to that of other fields. The physical quantity $\mathcal{X}$ has effect on the field potential $\mathcal{A}$, the angular momentum $\mathcal{S} \circ \mathcal{X}$, the energy $\mathcal{X}^* \circ (\mathcal{S} \circ \mathcal{X})$ and the power $(\mathcal{X} \circ \mathcal{X}^*) \circ (\mathcal{S} \circ \mathcal{X})$.

In the above equations, as a part of field potential $\mathcal{A}$, the item $(\mathcal{X}^* \circ \mathcal{X}/k)$ has an important impact on the subsequent equations. In the hyper-strong field, the force-balance equation can be obtained when $\mathcal{Z} = 0$, the conservation of angular momentum can be gained when $\mathcal{W} = 0$, and the energy conservation of can be attained when $\mathcal{N} = 0$.

Table 12.    Equations set of hyper-strong field

| Spacetime | Octonionic space H-S |
| --- | --- |
| $\mathcal{X}$ physical quantity | $\mathcal{X} = \mathcal{X}_{H-S}$ |
| Field potential | $\mathcal{A} = (\mathcal{X}/k + \diamond)^* \circ \mathcal{X}$ |
| Field strength | $\mathcal{B} = (\mathcal{X}/k + \diamond) \circ \mathcal{A}$ |
| Field source | $k\mu \mathcal{S} = k (\mathcal{X}/k + \diamond)^* \circ \mathcal{B}$ |
| Force | $\mathcal{Z} = k (\mathcal{X}/k + \diamond) \circ \mathcal{S}$ |
| Angular momentum | $\mathcal{M} = \mathcal{S} \circ (\mathcal{R} + k_{rx} \mathcal{X})$ |
| Energy | $\mathcal{W} = k (\mathcal{X}/k + \diamond)^* \circ \mathcal{M}$ |
| Power | $\mathcal{N} = k (\mathcal{X}/k + \diamond) \circ \mathcal{W}$ |

5.2 Quantum equations set of hyper-strong field

In the octonionic space H-S, the wave functions of the quantum mechanics are the octonionic equations set. And the Dirac and Klein-Gordon equations of the quantum mechanics are actually the wave equations set which are associated with particle's wave function $\Psi = \mathcal{M}/b$. Wherein, the wave function $\Psi = \mathcal{S} \circ (\mathcal{R} + k_{rx} \mathcal{X})/b$ can be written as the exponential form with the octonionic characteristics. The coefficient $b = \hbar_{H-S}$ is the Plank-like constant.

### 5.2.1 Equations set of Dirac and Klein-Gordon

By comparison, we find that the Dirac equation and Klein-Gordon equation can be attained respectively from the energy equation (10) and power equation (11) after substituting the operator k ($\mathcal{X}$/k +◇) for ($\mathcal{W}$/kb +◇).

The $\mathcal{U}$ equation of the quantum mechanics can be defined as
$$\mathcal{U} = (\mathcal{W}/k + b\diamond)^* \circ (\mathcal{M} / b) \qquad (12)$$

The $\mathcal{L}$ equation of the quantum mechanics can be defined as
$$\mathcal{L} = (\mathcal{W}/k + b\diamond) \circ (\mathcal{U} / b) \qquad (13)$$

In hyper-strong field, when the energy quantum equation $\mathcal{U} = 0$, the Dirac and Schrodinger equations can be deduced to describe the field source particle with spin 1/2. When the power quantum equation $\mathcal{L} = 0$, Klein-Gordon equation can be obtained to explain the field source particle with spin 0. [4-6]

### 5.2.2 Equations set of Dirac-like

By analogy with the above equations, the Dirac-like and the second Dirac-like equations can be procured from the Eqs.(5), (6), (7) and (8) respectively. Wherein, $\mathcal{X}$/b is the wave function also.

The $\mathcal{D}$ equation of the quantum mechanics can be defined as
$$\mathcal{D} = (\mathcal{W}/k + b\diamond)^* \circ (\mathcal{X} / b) \qquad (14)$$

The $\mathcal{G}$ equation of the quantum mechanics can be defined as
$$\mathcal{G} = (\mathcal{W}/k + b\diamond) \circ (\mathcal{D} / b) \qquad (15)$$

The $\mathcal{T}$ equation of the quantum mechanics can be defined as
$$\mathcal{T} = (\mathcal{W}/k + b\diamond)^* \circ (\mathcal{G} / b) \qquad (16)$$

The $O$ equation of the quantum mechanics can be defined as
$$O = (\mathcal{W}/k + b\diamond) \circ (\mathcal{T} / b) \qquad (17)$$

In the hyper-strong field, when the field source quantum equations $\mathcal{D} = 0$, $\mathcal{G} = 0$, $\mathcal{T} = 0$ and $O = 0$, the above Dirac-like equations can be deduced to describe the different characteristics of intermediate particles with different quantum quantities.

Table 13.   Quantum equations set of hyper-strong field

| | |
|---|---|
| Energy quantum | $\mathcal{U} = (\mathcal{W}/k + b\diamond)^* \circ (\mathcal{M}/b)$ |
| Power quantum | $\mathcal{L} = (\mathcal{W}/k + b\diamond) \circ (\mathcal{U}/b)$ |
| Field potential quantum | $\mathcal{D} = (\mathcal{W}/k + b\diamond)^* \circ (\mathcal{X}/b)$ |
| Field strength quantum | $\mathcal{G} = (\mathcal{W}/k + b\diamond) \circ (\mathcal{D}/b)$ |
| Field source quantum | $\mathcal{T} = (\mathcal{W}/k + b\diamond)^* \circ (\mathcal{G}/b)$ |
| Force quantum | $O = (\mathcal{W}/k + b\diamond) \circ (\mathcal{T}/b)$ |

### 5.3 Compounding particles

In hyper-strong field, the intermediate and field source particles can be tabled in Table 14. We can find the intermediate particles and other kinds of new and unknown particles which may be existed in the nature and expressed in parentheses.

Some particles may possess more than two sorts of general charges. If the field source particles possess S strong-charge, W weak-charge, E electric-charge, G gravitational-charge and general charges of hyper-strong field, they will take part in the strong interaction, weak interaction, electromagnetic interaction, gravitational interaction and hyper-strong field etc.

Table 14.   Sorts of particles in hyper-strong field

|  | A A-charge | B A-charge | A B-charge | B B-charge |
|---|---|---|---|---|
| A A-charge | (intermediate particle) | (green field source particle) | (red field source particle) | (blue field source particle) |
| B A-charge | (green field source particle) | (intermediate particle) | (?) | (?) |
| A B-charge | (red field source particle) | (?) | (intermediate particle) | (?) |
| B B-charge | (blue field source particle) | (?) | (?) | (quark, intermediate particle) |

## 6. Octonionic hyper-weak field

There is an analogy between the octonionic spacetime of the hyper-weak field and that of the hyper-weak field. According to the 'SpaceTime Equality Postulation', the spacetime, which is associated with the hyper-weak field and possesses the physics content, is adopted by the octonionic space (Octonionic space H-W, for short). In the octonionic space H-W, various characteristics of the hyper-weak field can be described uniformly and some equations set of the hyper-weak field can be obtained.

In the octonionic space H-W, the base $\mathcal{E}$ can be written as
$$\mathcal{E} = (\mathbf{1},\ \vec{i},\ \vec{j},\ \vec{k},\ \vec{e},\ \vec{I},\ \vec{J},\ \vec{K}) \tag{18}$$
and their multiplication are tabled as Table 10.

The displacement ( $r_0$, $r_1$, $r_2$, $r_3$, $R_0$, $R_1$, $R_2$, $R_3$ ) in the octonionic space is
$$\mathcal{R} = (r_0 + \vec{i}\, r_1 + \vec{j}\, r_2 + \vec{k}\, r_3) + (\vec{e}\, R_0 + \vec{I}\, R_1 + \vec{J}\, R_2 + \vec{K}\, R_3) \tag{19}$$
where, $r_0 = ct$, $R_0 = cT$. c is the speed of intermediate particle, t and T denote the time.

The octonionic differential operator $\Diamond$ and its conjugate operator $\Diamond^*$ are defined as,
$$\Diamond = \Diamond_{hw} + \Diamond_{HW}\ ,\quad \Diamond^* = \Diamond^*_{hw} + \Diamond^*_{HW} \tag{20}$$
where, $\Diamond_{hw} = \partial_{hw0} + \vec{i}\, \partial_{hw1} + \vec{j}\, \partial_{hw2} + \vec{k}\, \partial_{hw3}$, $\Diamond_{HW} = \vec{e}\, \partial_{HW0} + \vec{I}\, \partial_{HW1} + \vec{J}\, \partial_{HW2} + \vec{K}\, \partial_{HW3}$. $\partial_{hwj} = \partial/\partial r_j$, $\partial_{HWj} = \partial/\partial R_j$, j = 0, 1, 2, 3 .

In the hyper-weak field, the field potential ( $a_0$, $a_1$, $a_2$, $a_3$, $k_{hw} A_0$, $k_{hw} A_1$, $k_{hw} A_2$, $k_{hw} A_3$ ) is defined as
$$\mathcal{A} = \Diamond^* \circ \mathcal{X}$$
$$= (a_0 + \vec{i}\, a_1 + \vec{j}\, a_2 + \vec{k}\, a_3) + k_{hw} (\vec{e}\, A_0 + \vec{I}\, A_1 + \vec{J}\, A_2 + \vec{K}\, A_3) \tag{21}$$
where, $\mathcal{X} = \mathcal{X}_{H-W}$ is the physical quantity in octonionic space H-W; $k_{hw}$ is the coefficient.

6.1 Equations set of hyper-weak field

The hyper-weak field may be consisting of C and D interactions. In the hyper-weak field, there exist four types of subfields (C-C subfield, C-D subfield, D-C subfield and D-D subfield) and their field sources (C C-charge, C D-charge, D C-charge, D D-charge). The subfields of the hyper-weak field are hyper-long range in Table 15.

Table 15.   Subfield types of hyper-weak field

|  | C Interaction | D Interaction |
|---|---|---|
| C quaternionic space | C-C subfield<br>C C-charge<br>intermediate particle $\gamma_{CC}$ | D-C subfield<br>C D-charge<br>intermediate particle $\gamma_{DC}$ |
| D quaternionic space | C-D subfield<br>D C-charge<br>intermediate particle $\gamma_{CD}$ | D-D subfield<br>D D-charge<br>intermediate particle $\gamma_{DD}$ |

It can be predicted that the field strength of the C-C subfield is stronger than other three subfields. The field strength of the D-C subfield and C-D subfield may be equal, and both of them are weaker than that of D-D subfield. The physical features of each subfield of the hyper-weak field meet the requirement of Table 16.

In the hyper-weak field, the definition of the force is very different to that of other fields (hyper-strong, strong-weak or electromagnetic-gravitational fields), and thus the succeeding equations and operators should be revised. By analogy with the case of other fields, the octonionic differential operator $\diamondsuit$ in the hyper-weak field needs to be generalized to the new operator ($S/k + \diamondsuit$). That is because the field source is not as fundamental as the field potential or field strength, and the hyper-weak field belongs to hyper-long range. Therefore the physical characteristics of hyper-weak field can be studied from many aspects. Where, the coefficient $k = c = k^S_{H-W}$ is the constant.

The field strength $\mathcal{B}$ of the hyper-weak field can be defined as

$$\mathcal{B} = \diamondsuit \circ \mathcal{A} \tag{22}$$

then the field source and the force of the hyper-weak field can be defined respectively as

$$\mu S = \diamondsuit^* \circ \mathcal{B} \tag{23}$$

$$Z = k\,(S/k + \diamondsuit) \circ S \tag{24}$$

where, the mark (*) denotes octonionic conjugate. The coefficient $\mu$ is interaction intensity of the hyper-weak field.

The angular momentum of the hyper-weak field can be defined as ($k_{rx}$ is the coefficient)

$$\mathcal{M} = S \circ (\mathcal{R} + k_{rx}\,\mathcal{X}) \tag{25}$$

and the energy and power in the hyper-weak field can be defined respectively as

$$\mathcal{W} = k\,(S/k + \diamondsuit)^* \circ \mathcal{M} \tag{26}$$

$$\mathcal{N} = k\,(S/k + \diamondsuit) \circ \mathcal{W} \qquad (27)$$

The above equations show that, the angular momentum $\mathcal{M}$, energy $\mathcal{W}$ and power $\mathcal{N}$ of the hyper-strong field are different to that of other fields. The physical quantity $\mathcal{X}$ has effect on the field potential $\mathcal{A}$, the angular momentum $S \circ \mathcal{X}$, the energy $S^* \circ (S \circ \mathcal{X})$ and the power $(S \circ S^*) \circ (S \circ \mathcal{X})$.

In the above equations, as a part of force $\mathcal{Z}$, the item $(S \circ S/k)$ has an important impact on the subsequent equations. In hyper-weak field, the force-balance equation can be obtained when $\mathcal{Z} = 0$, the conservation of angular momentum can be gained when $\mathcal{W} = 0$, and the energy conservation can be attained when $\mathcal{N} = 0$.

Table 16.   Equations set of hyper-weak field

| Spacetime | Octonionic space H-W |
| --- | --- |
| $\mathcal{X}$ physical quantity | $\mathcal{X} = \mathcal{X}_{H\text{-}W}$ |
| Field potential | $\mathcal{A} = \diamondsuit^* \circ \mathcal{X}$ |
| Field strength | $\mathcal{B} = \diamondsuit \circ \mathcal{A}$ |
| Field source | $\mu S = \diamondsuit^* \circ \mathcal{B}$ |
| Force | $\mathcal{Z} = k\,(S/k + \diamondsuit) \circ S$ |
| Angular momentum | $\mathcal{M} = S \circ (\mathcal{R} + k_{rx}\,\mathcal{X})$ |
| Energy | $\mathcal{W} = k\,(S/k + \diamondsuit)^* \circ \mathcal{M}$ |
| Power | $\mathcal{N} = k\,(S/k + \diamondsuit) \circ \mathcal{W}$ |

6.2 Quantum equations set of hyper-weak field

In the octonionic space H-W, the wave functions of the quantum mechanics are the octonionic equations set. And the Dirac and Klein-Gordon equations of the quantum mechanics are actually the wave equations set which are associated with particle's wave function $\Psi = \mathcal{M}/b$. Wherein, the wave function $\Psi = S \circ (\mathcal{R} + k_{rx}\,\mathcal{X})/b$ can be written as the exponential form with the octonionic characteristics. The coefficient $b = \hbar_{H\text{-}W}$ is the Plank-like constant.

6.2.1 Equations set of Dirac and Klein-Gordon
By comparison, we find that the Dirac equation and Klein-Gordon equation can be attained respectively from Eqs.(24) and (25) after substituting operator $k\,(S/k + \diamondsuit)$ for $(\mathcal{W}/kb + \diamondsuit)$.

The $\mathcal{U}$ equation and $\mathcal{L}$ equation of the quantum mechanics can be defined respectively as

$$\mathcal{U} = (\mathcal{W}/k + b\diamondsuit)^* \circ (\mathcal{M}/b) \qquad (28)$$
$$\mathcal{L} = (\mathcal{W}/k + b\diamondsuit) \circ (\mathcal{U}/b) \qquad (29)$$

In the hyper-weak field, when the energy quantum equation $\mathcal{U} = 0$, Dirac and Schrodinger equations can be deduced to describe the field source particle with spin 1/2. When the power quantum equation $\mathcal{L} = 0$, the Klein-Gordon equation can be obtained to explain the field source particle with spin 0.

6.2.2 Equations set of Dirac-like
By analogy with the above equations, Dirac-like equation can be drawn from the force equation Eq.(24). Wherein, $\mathcal{S}$ /b is the wave function also.

The $\mathcal{S}$ equation of the quantum mechanics can be defined respectively as

$$O = (\mathcal{W}/k + b\diamondsuit) \circ (\mathcal{S}/b) \tag{30}$$

In the hyper-weak field, when the field source quantum equation $O = 0$, Dirac-like equation can be deduced to describe the intermediate particle.

Table 17.   Quantum equations set of hyper-weak field

| Energy quantum | $\mathcal{U} = (\mathcal{W}/k + b\diamondsuit)^* \circ (\mathcal{M}/b)$ |
|---|---|
| Power quantum | $\mathcal{L} = (\mathcal{W}/k + b\diamondsuit) \circ (\mathcal{U}/b)$ |
| Force quantum | $O = (\mathcal{W}/k + b\diamondsuit) \circ (\mathcal{S}/b)$ |

6.3 Compounding particles

In hyper-weak field, the intermediate and field source particles can be tabled as following. In Table 18, we can find the intermediate particles and other kinds of new and unknown particles which may be existed in the nature and expressed in parentheses. Some particles may possess more than two sorts of general charges. If the field source particles possess S strong-charge, W weak-charge, E electric-charge, G gravitational-charge, general charges of hyper-strong and hyper-weak fields, they will take part in the strong interaction, weak interaction, electro-magnetic interaction, gravitational interaction, hyper-strong and hyper-weak fields etc. [7-9]

Table 18.   Sorts of particles in hyper-weak field

|  | C C-charge | D C-charge | C D-charge | D D-charge |
|---|---|---|---|---|
| C C-charge | (intermediate particle) | (green field source particle) | (red field source particle) | (blue field source particle) |
| D C-charge | (green field source particle) | (intermediate particle) | (?) | (?) |
| C D-charge | (red field source particle) | (?) | (intermediate particle) | (?) |
| D D-charge | (blue field source particle) | (?) | (?) | (?, intermediate particle) |

7. Conclusions

By analogy with the octonionic electromagnetic-gravitational and strong-weak fields, the octonionic hyper-strong and hyper-weak fields have been developed, including their field equations, quantum equations and some new unknown particles.

In hyper-strong field, the study deduces Dirac equation, Schrodinger equation and Klein-Gordon equation of sub-quarks. And it is able to infer the Dirac-like and the second Dirac-like equations of intermediate particles among sub-quarks. It predicts that there may exist some new energy parts of hyper-strong field, new intermediate particles among the sub-quarks, and unknown particles which are associated with the sub-quark.

In the hyper-weak field, the research achieves the Yang-Mills equation, Dirac equation, Schrodinger equation and Klein-Gordon equation of the general charges in hyper-weak field. And it is able to conclude the Dirac-like and the second Dirac-like equation of intermediate bosons among the general charges of hyper-weak field. It predicts that there exist new energy components, some new unknown field source particles and the intermediate particles among general charges of the hyper-weak fielding in the galaxy clusters.

**Acknowledgements**

The author thanks Yun Zhu, Zhimin Chen and Minfeng Wang for helpful discussions. This project was supported by National Natural Science Foundation of China under grant number 60677039, Science & Technology Department of Fujian Province of China under grant number 2005HZ1020 and 2006H0092, and Xiamen Science & Technology Bureau of China under grant number 3502Z20055011.

**References**

[1] Zihua Weng. Octonionic electromagnetic and gravitational interactions and dark matter. arXiv: physics /0612102.

[2] Zihua Weng. Octonionic quantum interplays of dark matter and ordinary matter. arXiv: physics /0702019.

[3] Zihua Weng. Octonionic strong and weak interactions and their quantum equations. arXiv: physics /0702054.

[4] Ariel Zhitnitsky. Cold dark matter as compact composite objects. Phys. Rev. D, 74 (2006) 043515.
Yong-Jun Zhang, Wei-Zhen Deng, Bo-Qiang Ma. Principle of balance and the sea content of the proton. Phys. Rev. D, 65 (2002) 114005.

[5] K. Koike. A Schematic model of leptons and quarks based on Majorana partners. Prog. Theor. Phys., 108 (2002) 1165-1170.

[6] C. S. Kalman. Why quarks cannot be fundamental particles? Nucl. Phys. Proc. Suppl., 142 (2005) 235-237.

[7] Ilona K. Soechting, Mark E. Huber, Roger G. Clowes, Steve B. Howell. The FSVS cluster catalogue: galaxy clusters and groups in the faint sky variability survey. Mon. Not. Roy. Astron. Soc., 369 (2006) 1334-1350.

[8] Zackaria Chacko, Elena Perazzi. Extra dimensions at the weak scale and deviations from Newtonian gravity. Phys. Rev. D, 68 (2003) 115002.

[9] P. Barmby, K. D. Kuntz, J. P. Huchra, J. P. Brodie. Hubble space telescope observations of star clusters in M101. Astron. J., 132 (2006) 883.